\documentclass[12pt,a4paper]{article}

\begin{document}
\newtheorem{theorem}{Theorem}[subsection]
\newtheorem{lemma}[theorem]{Lemma}
\newtheorem{defn}{Definition}[subsection]
\newtheorem{ex}{Exercise}[subsection]
\renewcommand{\theequation}{\arabic{section}.\arabic{equation}}

\newcommand{\tightlist}[1]{\begin{list}{$\bullet$}
                        {\usecounter{enumi}\setlength{\parsep}{1pt}}
                          #1\end{list}}
\newcommand{\innerlist}[1]{\begin{list}{$\circ$}
                        {\usecounter{enumii}\setlength{\parsep}{1pt}
                          \setlength{\leftmargin}{8truept}}
                          #1\end{list}}
\def\sqr#1#2{{\vcenter{\vbox{\hrule height.#2pt
        \hbox{\vrule width.#2pt height#1pt \kern#1pt
           \vrule width.#2pt}
        \hrule height.#2pt}}}}
\def\qed{\begin{flushright}$\sqr84$\end{flushright}}
\newcommand{\bes}{\begin{eqnarray*}}
\newcommand{\ees}{\end{eqnarray*}}
\newcommand{\PI}{$\textrm{P}_{
       \textrm{\footnotesize I}}$}
\newcommand{\PII}{$\textrm{P}_{
       \textrm{\footnotesize II}}$}
\newcommand{\PIII}{$\textrm{P}_{
       \textrm{\small III}}$}
\newcommand{\PIV}{$\textrm{P}_{
       \textrm{\footnotesize IV}}$}
\newcommand{\PV}{$\textrm{P}_{
       \textrm{\small V}}$}
\newcommand{\PVI}{$\textrm{P}_{
       \textrm{\small VI}}$}
\newcommand{\PtIV}{$\textrm{P}_{
       \textrm{\tiny XXXIV}}$}
\newcommand{\PtV}{$\textrm{P}_{
       \textrm{\tiny XXXV}}$}

\newcommand{\Pa}{Painlev\'e}
\newcommand{\Ba}{B\"acklund}

\newcommand\cP{\mathcal{P}}
\title{Mappings preserving locations of movable poles:
a new extension of the truncation method to ordinary
differential equations}
\author{Pilar R.\ Gordoa\thanks{permanent address:
Facultad de Ciencias, Edificio de Fisica,
Universidad de Salamanca, 37008 Salamanca,
Spain,\ \texttt{prg@sonia.usal.es}}, 
Nalini Joshi\thanks{\texttt{njoshi@maths.adelaide.edu.au}} 
and Andrew Pickering\thanks{current address: Facultad de Ciencias, 
Edificio de Fisica,
Universidad de Salamanca, 37008 Salamanca,
Spain,\ 
\texttt{andrew@sonia.usal.es}}\\
        {\it Department of Pure Mathematics}\\
        {\it University of Adelaide}\\
        {\it Adelaide Australia 5005}}
\maketitle
\begin{center}
{\bf Abstract}
\end{center}
The truncation method is a collective name for techniques that arise
from truncating a Laurent series expansion (with leading term) of 
generic solutions
of nonlinear partial differential equations (PDEs). Despite its utility in
finding B\"acklund transformations and other remarkable properties 
of integrable PDEs, it has not been generally extended
to ordinary differential equations (ODEs). Here we give a new general 
method that provides such an extension and show how to apply it to
the classical nonlinear ODEs called the \Pa\ equations.
Our main new idea is to consider mappings that preserve the locations
of a natural subset of the movable poles admitted by the equation.
In this way we are able to recover all known fundamental \Ba\ 
transformations for the equations considered. We are also able to
derive \Ba\ transformations onto other ODEs in the \Pa\ classification.

\section[1]{Introduction}
Integrable differential equations are 
those that are solvable (for a large space of initial data)
through an associated linear problem\cite{ac:cup}. It is conjectured
that the solutions of all such equations possess 
a characteristic complex singularity
structure \cite{ars:jmp}. In particular, there
is widespread evidence that all movable singularities of all solutions 
are poles \cite{kjh:96,mdkpc:pp}. This is commonly referred
to as the \textit{\Pa\ property}. Extensions of this definition can be found
for example in \cite{cfp:neg}.

This property has been used as a starting point for deducing other
remarkable properties of integrable partial differential equations
(PDEs).  The basic idea, first proposed by Weiss\cite{weiss:II} (see
also \cite{wtc:jmp}), is to truncate a Laurent expansion of a generic
solution near a movable pole.  For example, the Korteweg-deVries
(KdV) equation 
\[ u_t+6uu_x+u_{xxx}=0,\quad u=u(x,t) \] 
admits locally
convergent Laurent series of the form \cite{njjap:nonlin,njgs:nonlin}
\begin{equation} 
u = {-\,2{\Phi_x}^2\over {\Phi}^2}+{2{\Phi_{xx}}\over
{\Phi}}+ \sum_{n=0}^{\infty}u_n(x,t){\Phi(x,t)}^n,
\label{laurent_kdv}
\end{equation} 
as solutions in a neighbourhood of any analytic,
noncharacteristic variety given by $\Phi(x,t)=0$. (Note that
noncharacteristic here implies $\Phi_x\not=0$.)  The \lq\lq
truncation\rq\rq\ of this series is 
\begin{equation} u =
{-\,2{\Phi_x}^2\over {\Phi}^2}+{2{\Phi_{xx}}\over {\Phi}} +g(x,t)
\label{trunc_kdv} 
\end{equation}                        
where $g(x,t)$ (often called the \lq\lq constant-level\rq\rq\ term)
is analytic near $\Phi=0$. Asking that this expression be a solution
of the KdV equation, order by order in $\Phi$, requires that $g$ must
be a solution of the KdV equation, and also that $\Phi$ satisfies an
equation, often referred to as the ``singular manifold equation.''
Weiss showed how it is then possible to deduce 
the well known linear problem and Darboux transformation for the KdV
equation. These then yield the \Ba\ transformation for the KdV equation.
Such ideas have led to a procedure called the		  
\lq\lq truncation method\rq\rq\ that has been successfully
extended \cite{egmr93,pilar1,mc:1,cmp:2,p:96,pilar2}
to many PDEs.		
				
However, no such general procedure exists for ODEs. The PDE-truncation
procedure relies on setting coefficients of different powers of
$\Phi$ to zero in the image equation obtained by substituting the
truncation.  This poses a difficulty for ODEs. In each case for which
this has been tried for an ODE \cite{w:84a,w:84b,ntz:87,gntz:88},
the results have been found to be very restricted. For example, only
special transformations or special solutions have been found via
this procedure. In particular, no general, parameter-dependent \Ba\
transformation has been found by a truncation method.

Recently, Clarkson, Joshi and Pickering \cite{cjp} have shown
how this difficulty can be overcome for the second \Pa\ (\PII )
hierarchy. The main idea here was to use the truncation and the
image equation (obtained by substituting the truncation) to eliminate
$\Phi$, instead of separating powers of $\Phi$. The result is a \Ba\
transformation for the hierarchy. This approach has also been extended
to the reductions of the modified Sawada-Kotera/Kaup-Kupershmidt
(mSK/mKK) hierarchy \cite{jp}. 

Our purpose here is to give a {\em universal}
truncation-type method for ODEs that is based on singularity
analysis. The above applications (to the \PII\ and reduced
mSK/mKK hierarchies) can be recast in this framework.  The main
new idea is to consider truncation as a mapping that preserves the
locations of a natural {\em subset} of movable singularities.

An example is given by the second \Pa\ equation		
\begin{equation}
y''=2y^3+xy+\alpha,\quad y=y(x),	\label{p2}
\end{equation}
whose general solution possesses two families of movable poles 
\cite{njmdk:direct}. (These are often referred to in the 
literature as two \lq\lq branches\rq\rq of a \Pa\ expansion.) 
Near a movable singularity $x=x_0$, say, $y(x)$ has a convergent 
Laurent expansion
\begin{equation}
y = {\pm\,1\over (x-x_0)}+h(x)(x-x_0),\label{laurent_p2}
\end{equation}
where $h(x)$ is locally analytic. Clearly, the set of poles of $y(x)$
naturally separates into two subsets, identified by the sign of
the coefficient of the leading-order pole. 

Each of the \Pa\ equations, except the first, has 
generic solution $y(x)$ that possesses pairs of simple movable poles
with coefficients of opposite sign.
Therefore, the set of all movable poles of a solution $y(x)$
decomposes into the union of two nonintersecting subsets $\cP_+$ and
$\cP_-$. By $\cP_+$ we mean the set of poles with positive choice of coefficient
and $\cP_-$ is that with the negative choice.

In the following section, a generic solution $y(x)$ of a \Pa\ equation
will be transformed
to a solution $Q(x)$
of the same equation but with possibly different parameters as
\begin{equation}
y(x)=\rho(x) + Q(x).\label{trunc}
\end{equation}
We construct the transformation by demanding
that $\rho(x)$ have poles exactly at the elements of
$\cP_+$ and $Q(x)$
have them at $\cP_-$ (or vice-versa).
When $Q(x)$ satisfies the same equation (albeit with different parameters),
we find B\"acklund transformations (BTs), through a procedure
that relies only on singularity analysis of the transformed equation.

The method is applied to \PII\ and \PIV\ in this paper. In section 2, 
we give details for \PII\ and indicate the major differences for \PIV .
This method also leads to related ODEs and \Ba\ transformations between 
these and the equation under consideration; these are considered in 
Section 3, again for \PII\ and \PIV .
In section 4, we show how to carry out the analogue of the so called
double singular manifold method, i.e. one which considers two
singularities of the solution simultaneously. Extensions of 	
these techniques to other Painlev\'e equations are considered in a 
second paper \cite{gjp2}.

\section{Truncation for \PII\ and \PIV }
In this section we show how to carry out a mapping that preserves
\lq\lq half\rq\rq\ the movable poles of the solution of a \Pa\ equation
and how this can lead to a \Ba\ transformation for that equation.

We recall the equations \PII\ and \PIV\  here for reference later.
\begin{eqnarray}
y''&=&2y^3+xy+\alpha\\
y''&=&{{y'}^2\over 2y}+{3 y^3\over
2}+4xy^2+2(x^2-\alpha)y-{\gamma^2\over 2y}
\end{eqnarray}
(Note the slightly unusual renaming of the second parameter in \PIV\
that differs from convention.)
We take $Q(x)$ to satisfy the same equation as $y(x)$ but with possibly
different parameters indicated by roman letters replacing the
corresponding greek ones. (So e.g. $\alpha\mapsto a$.)

\subsection{\PII }
Substitution of Eq (\ref{trunc}) into \PII\
gives
\begin{equation}
\rho'' - 2\rho^3-6\rho^2 Q - 6\rho Q^2 -x\rho -\alpha +a =0.
\label{rho_ode_p2}
\end{equation}
The dominant terms of this equation near a pole of $\rho$ are
\[
\rho'' \approx 2\rho^3\ \Rightarrow\ {\rho'}^2 \approx {\rho}^4.
\]
Taking the positive square root (i.e. taking $\rho$ to have
a pole in $\cP_-$), we write
\begin{equation}
\rho'(x)=:\rho(x)^2 + \sigma(x)\rho(x),\label{ricc_p2}
\end{equation}
where $\sigma$ is to be found. Using this to replace $\rho'$ in the
equation (\ref{rho_ode_p2}), we get
\[
\bigl(3\sigma(x)-6Q(x)\bigr)\rho(x)^2
+\left(-x+\sigma(x)^2+\sigma'(x) - 6 Q(x)^2\right)\rho(x)-\alpha+a=0.
\]

Consider the dominant terms of this equation near a pole of $\rho$.
Since $Q(x)$ is regular at such points, we get to leading-order
\[
\sigma(x)\approx 2Q(x)\ \Rightarrow\ \sigma(x)=:2Q(x) + {\tau(x)\over\rho(x)}
\]
Now Eq (\ref{rho_ode_p2}) becomes
\[
\left(-x+2Q'(x)-2Q(x)^2+2\tau(x)\right)\rho(x)+
2\tau(x)Q(x)+\tau'(x)+a-\alpha =0,
\]
from which we have
\[
\rho(x) = -\,{2\tau(x)Q(x)+\tau'(x)+a-\alpha\over -x+2Q'(x)-2Q(x)^2+2\tau(x)}.
\]

However, this must be compatible with the Riccati equation
(\ref{ricc_p2}) for $\rho(x)$. Substituting this expression for
$\rho(x)$ into (\ref{ricc_p2}) yields a compatibility condition
that contains $Q$, $Q'$ and $Q''$. We use the equation \PII\ satisfied
by $Q(x)$ to eliminate $Q''$. The result is a polynomial equation in
$Q$ and $Q'$, with coefficients involving $\tau$, $\tau'$ and $\tau''$,
given by
\begin{eqnarray}
0&=&  - 4\,\tau(x)Q(x)^{4} + 4\tau'(x)Q(x)^3\nonumber \\
&&+  \left(12{\tau}(x)^{2} + 12{ \tau}(x\,)Q'(x) + 2 \tau''(x) 
- 4x\,\tau(x)\right) \,{ Q}(\,{x}\,)^{2} \nonumber \\
&&  +\left(- 2\,{\tau}(x) +2 x \tau'(x) - 4 Q'(x)\tau'(x)  
+ 4a\,{ \tau}(x)\right)Q(x) \nonumber \\
&& - 8\,{ \tau}(x) {(Q'(x))}^{2} \nonumber\\
&&+  \left( - 2{\tau''(x)} + 6x \tau(x) - 12\,{ \tau}(x)^{2}\right)Q'(x)
 \nonumber \\
&&- {x}^{2}{ \tau}(x) +{a}^{2} - {\alpha}^{2} + { \alpha} -a+2a \tau'(x)\nonumber\\
&&- \tau'(x) +  \bigl(\tau'(x)\bigr)^{2} 
\nonumber \\
 & &- 2\,{ \tau''(x)}\,{ \tau}(\,{x}\,) - 4\,{ \tau}(\,{x}
\,)^{3}  + x\,{\tau''(x)} + 4x\,{ \tau}(\,{x}\,
)^{2}\label{p2_compat}
\end{eqnarray}
Since the solution space of \PII\ depends on two arbitrary parameters given by
$Q$, $Q'$ at a (regular) point, we can demand that this equation be
satisfied identically in these two variables. Since
$\rho$ is a functional of $Q$ (or equivalently $y$), $\tau$ is also.
The simplest solutions of Eq (\ref{p2_compat}) are those 
that are polynomial in $Q$ and $Q'$. Inspection of Eq (\ref{p2_compat})
shows that the simplest solution independent of both $Q$ and $Q'$ is 
$\tau(x)\equiv 0$ (because of the presence of the
monomial $4\tau(x)Q(x)^4$) under the constraint
\[
a^2-a = \alpha^2-\alpha\ \Rightarrow\ a=\alpha\ \textrm{or}\ a =-\alpha +1 .
\]
Then, the above solution for $\rho$ becomes 			
\begin{equation}
\rho(x) = {a-\alpha\over x-2Q'(x)+2Q(x)^2} . 
\label{bt1}
\end{equation}
Thus in addition to the identity $y(x)=Q(x)$ we obtain 
the well known BT for \PII\ \cite{gambier:acta}.
 
We have also considered other possible 
solutions for $\tau$, e.g. those that only
depend polynomially on $Q$ but not $Q'$. The monomial $- 2{\tau''(x)}Q'(x)$
in Eq (\ref{p2_compat}) shows that it can be at most linear in $Q$. 
Other monomials in the resulting equation then 
lead again to $\tau(x)\equiv 0$. 
(For other equations, a possible dependence of $\tau$ on $Q$ can give
more general results \cite{gjp2}.)

If in Eq (\ref{ricc_p2}) we had taken the negative root, i.e.
\[
\rho'(x)=-\rho(x)^2 + \sigma(x)\rho(x),
\] 
the above procedure would have yielded		
\[
\rho(x) = {a-\alpha\over x+2Q'(x)+2Q(x)^2}, \quad a=-\alpha-1 ,
\]
the alternative form of the BT for \PII . This could also have
been found by using the discrete symmetry of \PII\  under
$y\mapsto -y$, $\alpha\mapsto -\alpha$ in combination with 
Eqn (\ref{bt1}). It should be noted, however, that both BTs
are needed in order to iterate to find sequences of special 
integrals and rational solutions.

\subsection{\PIV }
Here we describe major differences in applying the above procedure
to \PIV . The first difference lies in the dominant
terms of the transformed equation. These are 
\[
\rho''\rho-{{\rho'}^2\over 2}\approx {3\rho^4\over 2}.
\]
After using the integrating factor $\rho'\rho^{-2}$, we can
integrate to get
\[
{\rho'^2\over \rho}\approx \rho^{3}\ \Rightarrow\ 
\rho'\approx\pm\rho^2.
\]
Consider first the case with minus sign on the right, i.e.
\[
\rho'=-\rho^2+\sigma \rho.
\]
Then we find (following dominant balances again) that
\[
\sigma(x)=-2x-2Q(x)+{\tau(x)\over\rho(x)}
\]
Using this in the transformed equation leads to a 
quadratic equation for $\rho$:
\begin{eqnarray}
& &\Biggl( -\, 2x Q(x) -
{ \tau}(x) -  Q({x})^{2} - 2 + 2\,{ \alpha} -
Q'({x})
\Biggr) \rho({x})^{2} +\nonumber \\
 & &\quad\Biggl(4\alpha Q({x}) - {\displaystyle \frac {1}{2
}} Q({x})^{3} - 2\, Q({x})\,{a} -
{\displaystyle \frac {1}{2}}\,{\displaystyle \frac {{c}^{2}}{
Q({x})}} + {\displaystyle \frac {1}{2}}\,
{\displaystyle \frac {
{Q'({x})}^{2}}{ Q({x})
}} \nonumber\\
 & &\quad +  \tau'(x)
+ 2\,{x}^{2}\, Q({x})
 - 2\,\tau({x})\, Q({x}) - 2\, Q({x}) +
2\, Q'(x)
\,{x} \Biggr)\rho(x)\nonumber\\
& &\quad\quad - 2\, Q({x})\,{ \tau}(x)
{x} - 2\, Q({x})^{2}\,{a} - {\displaystyle \frac {1
}{2}}\,{ \tau}(\,{x}\,)^{2} + {\displaystyle \frac {1}{2}}\,{
\gamma}^{2} -
 Q'({x}) \,{ \tau}(\,{x}\,)\nonumber \\
 & &\quad\quad + Q({x})\,
\tau'({x})  + 2\,{ \alpha
}\,Q({x})^{2} - 2\,{ \tau}(\,{x}\,)\, Q({x})^{
2} - {\displaystyle \frac {1}{2}}\,{c}^{2}=0.\label{p4minus}
\end{eqnarray}
That we obtain a quadratic equation in $\rho$ is a second difference
between our results for \PII\ and \PIV . In general  (beginning with
a polynomial ODE that gives a Riccati equation for $\rho$) we could 	
obtain at this stage a polynomial of higher degree in $\rho$.

This quadratic equation must be compatible with the differential
equation
\begin{equation}
\rho'(x)=-\rho(x)^2-(2x+2Q(x))\rho(x)+{\tau(x)},
\label{p4ric}
\end{equation}
for $\rho(x)$. Since equation (\ref{p4minus}) is quadratic in $\rho$,
it is worth noting how we obtain a unique $\rho$ to check the
compatibility condition. Suppose we differentiate Eqn (\ref{p4minus}) w.r.t.
$x$ and use Eqn (\ref{p4ric}) to replace $\rho'$. Then we obtain a cubic 
equation in $\rho$. The third degree term can be eliminated
by using Eqn (\ref{p4minus}) multiplied by $\rho$. 
This yields another quadratic equation
for $\rho$. Eliminating the second-degree term by using Eqn (\ref{p4minus})
again, we get a linear equation
for $\rho$, which we solve. The result is substituted into 
Eqn (\ref{p4ric}) to obtain the compatibility condition we investigate.

This equation can be analysed to find $\tau$ and any conditions on the 
parameters. We assume here that $\tau$ is  		
independent of both $Q$ and $Q'$; the final results are the same 
if we allow $\tau$ to depend on $Q$. The coefficient of $Q^{13}$ gives
\begin{equation}
\tau(x)={2\over 3}\,(\alpha - a).
\label{tdag}
\end{equation}
Substitution of this into the compatibility condition yields
\[
c^2=\gamma^2+{4\over 3}\,(\alpha^2-a^2-2\alpha + 2a).
\]
With this choice of $c^2$, the remaining terms in the compatibility 
condition factor to give
\[
(\alpha -a)(2\alpha -6-3\gamma+4a)(2\alpha-6+3\gamma+4a)=0.
\]
Thus in addition to the identity $y(x)=Q(x)$ we obtain the two nontrivial 
BTs
\begin{equation}
\rho(x) = \,{(\pm\gamma-2+2\alpha)Q(x)\over 
Q'(x)+{Q(x)}^2+2xQ(x)\pm\gamma/2+1-\alpha}
\label{rp41}
\end{equation}
\begin{eqnarray}
a&=&-{1\over 4}\,(2\alpha - 6 \pm 3\gamma)
\label{daga} \\
c^2&=&(\alpha \mp \gamma/2-1)^2
\label{dagc}
\end{eqnarray}
where the choice of sign of $\gamma$ (and $c$) arises because of the
way we have written the second parameter in \PIV . This is the BT
labelled by \lq\lq dagger\rq\rq\ ($y^{\dagger}$) in \cite{bch:95}.

The alternative choice of sign in the dominant balance of terms in $\rho$ 
leads to the Riccati equation
\[
\rho'(x)={\rho(x)}^2+\sigma(x)\rho(x),
\]
with
\[
\sigma(x)=2Q(x)+2x+\tau(x)/\rho(x),
\]
where
\[
\tau(x) = {2\over 3}\,(a-\alpha).
\]
This leads to the two BTs
\[
\rho(x) = {(\pm\gamma-2-2\alpha)Q(x)\over
\,Q'(x)-{Q(x)}^2-2xQ(x)\pm\gamma/2+1+\alpha}
\]
\bes
a&=&-{1\over 4}\,(2\alpha + 6 \mp 3\gamma)\\
c^2&=&(\alpha \pm \gamma/2+1)^2
\ees
This is the BT labelled by \lq\lq double dagger\rq\rq ($y^{\ddagger}$)
in \cite{bch:95}. 

It is interesting to note that all the known BTs \cite{bch:95} of \PIV\ can 
be expressed in terms of only two fundamental BTs. In \cite{bch:95}, these 
two BTs are labelled by \lq\lq hat\rq\rq ($\hat{y}$) and \lq\lq tilde\rq\rq 
($\widetilde{y}$). In the next section, we show that our derivation of the 
dagger and double dagger BTs also gives rise to these two fundamental BTs.

\section{Related BTs: the ODEs satisfied by {$\mathbf \rho(x)$}}

The results of the previous Section were obtained by searching for auto-\Ba\ 
transformations. That is, we looked for transformations between two copies
of the same equation distinguished by possibly different parameter values.
In this Section, we consider the ODE satisfied by the function $\rho(x)$
in (\ref{trunc}), and deduce the \Ba\ transformation between this ODE and
the \Pa\ equation. 

\subsection{\PII }

In constructing our first auto-BT for \PII (see the previous 
Section), we obtained a solution $\rho$ of the Riccati equation
\[
\rho'(x) = \rho^2+2Q(x)\rho(x)
\]
given by	
\[
\rho(x) = {1-2\alpha \over 2{Q(x)}^2-2Q'(x)+x},
\]
where $Q(x)$ satisfies \PII\ with parameter $a=-\alpha+1$. Now we 
eliminate $Q(x)$ from the above equations, to find a second
order ODE satisfied by $\rho(x)$, together with \Ba\ transformations to
\PII\ (in both $y(x)$ and $Q(x)$).

Eliminating $Q(x)$ between the above two equations gives 
\[
\rho''(x) = \frac{3}{2}{\rho'(x)^2 \over \rho(x)} +\frac{1}{2} \rho(x)^3
            +x\rho(x)-(1-2\alpha)
\]
which under $\rho(x)=1/s(x)$ becomes
\[
s''(x) = \frac{1}{2}{s'(x)^2 \over s(x)} +(1-2\alpha)s(x)^2 -xs(x)
        -\frac{1}{2s(x)}
\]
which is \PtIV\ (i.e. the thirty-fourth equation in the classification
results presented in Chapter 14 of \cite{ince}). Using this change of 
variables, and the expression for $Q$
in terms of $\rho$, the BT (\ref{trunc}) becomes 	
\[
y(x) = {1-s'(x) \over 2s(x)}.
\]
Therefore, we recover the well-known mapping between \PtIV\ and \PII . We
also have, of course, a \Ba\ transformation from \PtIV\ to \PII\ in $Q(x)$.
If we had started with the second \Ba\ transformation for \PII , we again find
(after a simple change of variables) the mapping to \PtIV .  

\subsection{\PIV }

We now consider the ODE satisfied by $\rho(x)$ in our construction of 
\Ba\ transformations for \PIV . Recall from Equations (\ref{p4ric}), 
(\ref{tdag}) and (\ref{daga}), that $\rho(x)$ satisfies the Riccati 
equation
\begin{equation}
\rho'(x) = -\rho^2-2(Q(x)+x)\rho(x) +(\alpha-1\pm\gamma/2),
\label{dagRic}
\end{equation}
for the dagger transformation. Elimination of $Q(x)$ between this and
(\ref{rp41}) gives, after the substitution $\rho(x)=\pm C/s(x)$, the ODE
\begin{equation}
s''(x) = \frac{1}{2} {s'(x)^2 \over s(x)} +\frac{3}{2} s(x)^3 +4xs(x)^2
        +2(x^2-A) s(x) -\frac{C^2}{2s(x)}
\label{P4a}
\end{equation}
where
\bes
A & = & -\frac{1}{2} -\frac{1}{2}\alpha \pm\frac{3}{4}\gamma \\
\pm C & = & 1 -\alpha \mp\frac{1}{2}\gamma.
\ees
This is another copy of the same equation, \PIV . The corresponding 
auto-\Ba\ transformation is obtained from (\ref{trunc}) after 
eliminating $Q(x)$ by using Equation (\ref{dagRic}). The result is
\[
y(x) = {s'(x)-s(x)^2-2xs(x)+(1 -\alpha \mp\gamma/2) \over 2s(x)}.
\]
This is the BT labelled by \lq\lq tilde\rq\rq\ ($\widetilde{y}$) in \cite{bch:95}.

On the other hand, Equation (\ref{rp41}) gives
\[
s(x) = -{Q'(x)+Q(x)^2+2xQ(x)+(1+A\pm C/2) \over 2Q(x)}
\]
where by replacing $\alpha$, $\gamma$ in terms of $A$, $C$ above, we get
the parameters $a$, $c$ of the version of \PIV\ satisfied by $Q(x)$ as
\bes
a & = & \frac{1}{2} -\frac{1}{2}A \pm\frac{3}{4}C \\
c^2 & = & \frac{1}{4}\left(2+2A\pm C\right)^2.
\ees
Hence we get another BT for \PIV .
This is the BT labelled by \lq\lq hat\rq\rq\ ($\hat{y}$) in \cite{bch:95}. That is,
we have used our derivation of $y^{\dagger}$ to deduce both $\widetilde{y}$ 
and $\hat{y}$. 

The BTs $\widetilde{y}$ and $\hat{y}$ are fundamental BTs in the sense that all 
known BTs for \PIV\ can be expressed in terms of these two. (See \cite{bch:95}.)
Our approach shows that $y^{\dagger}$ which maps $Q(x)$ to $y(x)$ can be
considered, with appropriate choices of signs, as the composition 
$\widetilde{y}\circ\hat{y}$.

Now consider our derivation of the double dagger BT in Section 2.2. Following 
the same procedure of elimination of $Q(x)$, we find (with the substitution 
$\rho=\mp C/s(x)$) the same \PIV\ i.e.\ Equation (\ref{P4a}), but with
$A$ and $C$ given now by
\bes
A & = & \frac{1}{2} -\frac{1}{2}\alpha \mp\frac{3}{4}\gamma \\
\pm C & = & 1 +\alpha \mp\frac{1}{2}\gamma.
\ees
The corresponding BT for \PIV\ is
\[
y(x) = -{s'(x)+s(x)^2+2xs(x)+(1+\alpha\mp\gamma/2) \over 2s(x)}.
\]
This is $\hat{y}$. Elimination of $y(x)$ instead gives
\[
s(x) = {Q'(x)-Q(x)^2-2xQ(x)+(1-A\pm C/2) \over 2Q(x)}
\]
where 
\bes
a & = & -\frac{1}{2} -\frac{1}{2}A \mp\frac{3}{4}C \\
c^2 & = & \frac{1}{4}\left(2-2A\pm C\right)^2,
\ees
which is $\widetilde{y}$. Thus our derivation of $y^{\ddagger}$ also
allows us to obtain the fundamental BTs $\widetilde{y}$ and $\hat{y}$. Our
approach then shows that, for suitable choices of signs, $y^{\ddagger}$ 
can be expressed in terms of $\widetilde{y}$ and $\hat{y}$ as the composition 
$\hat{y}\circ\widetilde{y}$ \cite{bch:95}.

\section{The double-singularity approach}
In Sections 2 and 3, we assumed that $\rho$ inherited half the
poles of the \Pa\ transcendent $y(x)$. Now we consider the possibility that
both families of movable poles of $y(x)$ are inherited by specified
functions called respectively $\rho_1$ and $\rho_2$ in the transformation. 
That is, we rewrite $y$ as
\begin{equation}
y(x) = \rho_1(x) - \rho_2(x) + Y(x),
\label{double}
\end{equation}
where we assume that $Y(x)$ satisfies \PII , or respectively \PIV , 
with different parameters ($a$, or respectively $a$ and $c$). 

As before, dominant terms 
of \PII\ or \PIV\ lead to Riccati equations for $\rho_i$ to leading-order 
near a movable pole. The dominant terms in each equation are as before.
However, the lower-order terms cannot be uniquely 	
determined by dominant balances. For simplicity, we choose the Riccati
equation in $\rho_1$, to be linear in $\rho_2$ and {\it vice versa}\/. 

Consider the Riccati equations for $\rho_i$
to have the form
\begin{eqnarray}
{\rho_1}'(x)&=& {\rho_1(x)}^2+A_1(x)\rho_1(x)\rho_2(x)+B_1(x)\rho_1(x)
	\nonumber\\
 & &\quad +C_1(x)\rho_2(x)+\tau_1(x)\label{r1}\\
{\rho_2}'(x)&=& {\rho_2(x)}^2+A_2(x)\rho_1(x)\rho_2(x)+B_2(x)\rho_1(x)
         \nonumber\\
 & &\quad +C_2(x)\rho_2(x)+\tau_2(x)\label{r2}
\end{eqnarray}
In this case, where we take the same sign against $\rho_i^2$ in each Riccati
equation, this approach is analagous to the double singular manifold
method. However, we show at the end of each subsection below 	
that we can also obtain nontrivial results
by taking opposite signs in the above Riccati equations.

\subsection{\PII }
We transform \PII\ by using the relation (\ref{double}) and equations
(\ref{r1}), (\ref{r2}). Then we get
\begin{eqnarray}
& &- { \alpha} +a-{C}_{2}(x){ \tau}_{2}(x)+{\tau_1}'(x)
+{B}_{1}(x){ \tau}_{1}(x)-{B}_{2}(x){ \tau}_{1}(x)\nonumber\\
& &\quad + {C}_{1}(x){ \tau}_{2}(x)-{\tau_2}'(x)\nonumber\\
& &\quad + \Bigl(3B_1(x)-B_2(x)-6Y(x)-A_2(x)B_2(x)+A_1(x)B_2(x)\Bigr)
  {\rho_1(x)}^2\nonumber\\
& &\quad + \Bigl({B_1}'(x)-B_2(x)B_1(x)+C_1(x)B_2(x)-{B_2}'(x)\nonumber\\
& &\quad\quad -6{Y(x)}^2 + A_1(x){ \tau}_{2}(x)+{B_1(x)}^2-C_2(x)B_2(x)
  +2\tau_1(x)-x\nonumber\\
& &\quad\quad -A_2(x)\tau_2(x)\Bigr)\rho_1(x)+\Bigl(C_1(x)C_2(x)+
  A_1(x){ \tau}_{1}(x)+6{Y(x)}^2 \nonumber\\
& &\quad\quad +B_1(x)C_1(x) -{C_2(x)}^2+x +{C_1}'(x)-{C_2}'(x)-2\tau_2(x)\nonumber\\
& &\quad\quad -C_1(x)B_2(x)-A_2(x)\tau_1(x)\Bigr)\rho_2(x)\nonumber\\
& &\quad +\Bigl(C_1(x)-6Y(x)-3C_2(x)-C_1(x)A_2(x)+A_1(x)C_1(x)\Bigr)
	{\rho_2(x)}^2
 \nonumber\\
& &\quad +\Bigl(-A_2(x)+A_1(x)A_2(x)+3A_1(x)+6-{A_2(x)}^2\Bigr)
 {\rho_1(x)}^2\rho_2(x)\nonumber\\
& &\quad +\Bigl(A_1(x)-A_1(x)A_2(x)-3A_2(x)-6+{A_1(x)}^2\Bigr)
 \rho_1(x){\rho_2(x)}^2\nonumber\\
& &\quad +\Bigl(2C_1(x)-{A_2}'(x)-2B_2(x)+A_1(x)C_2(x)+C_1(x)A_2(x)\nonumber\\
& &\quad \quad +12 Y(x)-2C_2(x)A_2(x)+2B_1(x)A_1(x)-A_1(x)B_2(x)\nonumber\\
& &\quad \quad +{A_1}'(x)-A_2(x)B_1(x)\Bigr)\rho_1(x)\rho_2(x)=0.\label{p2_double}
\end{eqnarray}

The dominant balances of this equation give
\begin{eqnarray}
-A_2(x)+A_1(x)A_2(x)+3A_1(x)+6-{A_2(x)}^2&=&0\label{112}\\
3B_1(x)-B_2(x)-6Y(x)-A_2(x)B_2(x)+A_1(x)B_2(x)&=&0\label{11}\\
A_1(x)-A_1(x)A_2(x)-3A_2(x)-6+{A_1(x)}^2&=&0\label{122}\\
C_1(x)-6Y(x)-3C_2(x)-C_1(x)A_2(x)+A_1(x)C_1(x)&=&0\label{22}
\end{eqnarray}
The sum of Eqns (\ref{112}) and (\ref{122}) factors to give
\[
(A_2-A_1)(A_1+A_2+4)=0.
\]
Substitution of the first solution $A_2=A_1$ shows that 
\begin{equation}
A_2 = A_1 = -3
\label{p2_case1}
\end{equation}
On the other hand, the second solution $A_1=-A_2-4$ 
gives
\[
A_1=-1, A_2=-3\ \textrm{or}\ A_1 = -3, A_2 = -1.  
\]
By relabelling if necessary, we take this case to be
\begin{equation}
A_1= -1, A_2= -3.
\label{p2_case2}
\end{equation}
\subsubsection{Case 1: $A_2 = A_1 = -3$}
Consider the first case (\ref{p2_case1}). Equations 
(\ref{11}) and (\ref{22}) give
\[
3B_1=B_2+6Y,\quad C_1=3C_2+6Y.
\]
We use these to replace $B_1$, $C_1$ in the transformed
equation (\ref{p2_double}). Now the dominant terms give
\begin{eqnarray}
3\tau_1&=&{B_2}'(x)+{3x\over 2}+{{B_2(x)}^2\over 3}-8B_2(x)Y(x)+3Y(x)^2\nonumber\\
&&\quad -3C_2(x)B_2(x)-3Y'(x)\label{c11}\\
\tau_2&=&{C_2}'(x)+{x\over 2}+{C_2(x)}^2-2B_2(x)Y(x)+9Y(x)^2\nonumber\\
&&\quad +6C_2(x)Y(x)+3Y'(x)\label{c12}-C_2(x)B_2(x)
\end{eqnarray}
The transformed equation then becomes
\bes
 & & -\alpha-3a+\Bigl(C_2(x)-{\displaystyle\frac{1}{3}}B_2(x)\Bigr)x
-2C_2(x)^2 B_2(x) +2C_2(x)^3 -{\displaystyle\frac{2}{27}} B_2(x)^3
\nonumber \\
& & +{\displaystyle\frac{2}{3}} C_2(x) B_2(x)^2 +2 B_2(x)^2 Y(x) +18
C_2(x)^2 Y(x) -12 C_2(x) B_2(x) Y(x) 
\nonumber \\
& & +54 C_2(x) Y(x)^2 -18 B_2(x) Y(x)^2 +{\displaystyle\frac{1}{3}} 
B_2''(x) - C_2''(x) = 0,
\ees
which, under the change of variables
\[
B_2(x)=3C_2(x)+3V(x)+9Y(x)
\]
becomes
\[
V''(x)-2V(x)^3-xV(x)-\alpha=0,
\]
which is just \PII\ with parameter $\alpha$ (the same parameter as
that in the version of \PII\ satisfied by $y(x)$). 	

These results for the coefficients $B_1$, $C_1$, $B_2$, motivate 	
the change of variables
\bes
\rho_1(x)&=&C_2(x)+2Y(x)+\sigma_1(x) \\
\rho_2(x)&=&C_2(x)+3Y(x)+V(x)+\sigma_2(x)
\ees
which simplify the Riccati equations for $\rho_1$ and $\rho_2$. These become
\begin{eqnarray}
{\sigma_1}'(x)&=&{\sigma_1(x)}^2-3\sigma_1(x)\sigma_2(x)\nonumber\\
& &\quad -2\sigma_1(x)V(x)+V'(x)+V(x)^2+x/2\label{sig1}\\
{\sigma_2}'(x)&=&{\sigma_2(x)}^2-3\sigma_1(x)\sigma_2(x)\nonumber\\
& &\quad +2\sigma_2(x)V(x)-V'(x)+V(x)^2+x/2\label{sig2}
\end{eqnarray}

We can eliminate $\sigma_2(x)$ from this system by solving the first
equation for $\sigma_2(x)$ and substituting the result into the second.
The result is, for $W(x)=2\sigma_1(x)$, 
\begin{eqnarray}
W''(x)&=&{2\over 3}\,{{W'(x)}^2\over W(x)}-{\displaystyle\frac{1}{3}}
\left(2W(x)-2V(x)+{2V'(x)+x+2V(x)^2\over W(x)}\right)W'(x)\nonumber\\
& &\quad +{\displaystyle\frac{2}{3}}{W(x)}^3
         -{\displaystyle\frac{10}{3}}V(x){W(x)}^2
+{\displaystyle\frac{1}{3}}\left(6V(x)^2+10V'(x)-x\right)W(x)\nonumber\\
& &\quad +{\displaystyle\frac{8}{3}}V(x)^3
         +{\displaystyle\frac{4}{3}}xV(x)+1
         +{\displaystyle\frac{8}{3}}V(x)V'(x)+2\alpha
         -{\displaystyle\frac{1}{3}}\left(x^2+4V(x)^4 \right.\nonumber\\
& &\quad \left. +4xV'(x)+8V(x)^2V'(x)
         +4xV(x)^2+4{\bigl(V'(x)\bigr)}^2\right){1\over W(x)}\label{w35}
\end{eqnarray}

This is \PtV\ whose general form (where the functions $r(x)$ and $q(x)$
are as given in \cite{ince}) is      
\bes
{d^2W\over dx^2}&=&{2\over 3W}\left({dW\over dx}\right)^2
 -\left({2W\over 3}-{2q\over 3}-{r\over W}\right){dW\over dx}\\
&&\quad +{2W^3\over 3}-{10\over 3}qW^2+\left(4q'+r+8q^2/3\right)W\\
&&\quad +2qr-3r'-{3r^2\over W}.
\ees
Equation (\ref{w35}) is this equation
with choices of coefficients given by
\[
q(x)=V(x),\quad r(x)=-{2\over 3}\Bigl(V'(x)+V(x)^2\Bigr)-\frac{1}{3}x .
\]
(We could have eliminated $\sigma_1$ in the above instead of $\sigma_2$.
The result is the same except for a sign change: $q=-V$.)
Writing the \Ba\ transformation in terms of $W(x)=2\sigma_1(x)$ gives
\bes
y(x)&=& \sigma_1(x)-\sigma_2(x)-V(x)\\
    &=& {\displaystyle\frac{1}{3}}W(x)
       -{\displaystyle\frac{1}{3}}{x \over W(x)}
       +{\displaystyle\frac{1}{3}}{W'(x) \over W(x)}
       -{\displaystyle\frac{2}{3}}{V(x)^2 \over W(x)}
       -{\displaystyle\frac{1}{3}} V(x)
       -{\displaystyle\frac{2}{3}}{V'(x) \over W(x)}
\ees
with inverse
\[
W(x) = {y'(x)-V'(x)+y(x)^2-V(x)^2 \over y(x)-V(x)} . 
\]
Hence we have reproduced the relation between two solutions of \PII\
and a solution of \PtV\ that has been known since classical studies
of these equations \cite{ince}.  

\subsubsection{Case 2: $A_1= -1, A_2= -3.$}
Now we consider the second case of possible values of $A_1$, $A_2$ 
given by Eqn (\ref{p2_case2}).                    
Here we find from (\ref{11}) and (\ref{22}) that
\bes
C_1(x) & = & C_2(x)+2Y(x) \\
B_1(x) & = & -{\displaystyle\frac{1}{3}}B_2(x)+2Y(x) .
\ees
Substituting these values of $C_1(x)$ and $B_1(x)$ into
(\ref{p2_double}) gives (from the coefficient of $\rho_1(x)\rho_2(x)$)
\[
B_2(x) = 3C_2(x)+9Y(x).
\]
The transformed equation then gives 
\bes
\tau_1(x) & = & C_2'(x)-C_2(x)Y(x)-2Y(x)^2+2Y'(x) \\
\tau_2(x) & = & C_2'(x)-2C_2(x)^2-9C_2(x)Y(x)-9Y(x)^2+3Y'(x)+\frac{1}{2}x
\ees
and substitution of these values of $\tau_1(x)$ and $\tau_2(x)$ gives
\[
\alpha = -\,1/2.
\]
Our subsequent results apply only to this special case of \PII . 

Making the change of variables
\bes
\rho_1(x)&=&C_2(x)+2Y(x)+\sigma_1(x)\\
\rho_2(x)&=&C_2(x)+3Y(x)+\sigma_2(x)
\ees
we find that the BT becomes 
\[
y(x)=\sigma_1(x) - \sigma_2(x).
\]
Note that $Y$ and $a$ have been eliminated here.
The equations satisfied by $\sigma_1(x)$ and $\sigma_2(x)$ now yield
two (apparently) different second-order equations. Elimination
of $\sigma_2(x)$ gives 
\[
\sigma_1''(x)-2\sigma_1(x)^3+\frac{1}{2}x\sigma_1(x)=0
\]
which is equivalent to \PII\ with zero value of the parameter. The BT in 
this case is
\[
y(x)={{\sigma_1}'(x)\over \sigma_1(x)},
\]
where $y(x)$ satisfies \PII\ with $\alpha = -\,1/2$. This BT, from \PII\ 
with $\alpha = 0$ to \PII\ with $\alpha = -\,1/2$, does not seem to be
widely known, although it can in fact be found in \cite{gambier:acta}.

On the other hand, elimination of $\sigma_1(x)$ gives (with 
$\sigma_2(x)=-W(x)$)
\begin{eqnarray}
W''(x) & = & \frac{2}{3}{(W'(x))^2 \over W(x)} -\frac{1}{3}
\left(2W(x)-{x \over 2W(x)} \right)W'(x) +\frac{2}{3} W(x)^3 
\nonumber\\
& & +\frac{1}{6}xW(x)-\frac{1}{12} {x^2 \over W(x)} -\frac{1}{2}
\label{ff1}
\end{eqnarray}
which can be rescaled onto \PtV\ with $q=0$, $r=-x/3$. The BT in this 
case is
\begin{equation}
y(x)=  {2\over 3}W(x) - {W'(x)\over 3W(x)} -{x\over 6W(x)}.
\label{ff2}
\end{equation}

We now briefly consider the results obtained from our double-singularity 
approach if we assume our Riccati system to be
\begin{eqnarray}
{\rho_1}'(x)&=& {\rho_1(x)}^2+A_1(x)\rho_1(x)\rho_2(x)+B_1(x)\rho_1(x)
	\nonumber\label{rr1}\\
 & &\quad +C_1(x)\rho_2(x)+\tau_1(x)\\
{\rho_2}'(x)&=& -{\rho_2(x)}^2+A_2(x)\rho_1(x)\rho_2(x)+B_2(x)\rho_1(x)
         \nonumber\label{rr2}\\
 & &\quad +C_2(x)\rho_2(x)+\tau_2(x)
\end{eqnarray}
With this choice of Riccati system we obtain two inequivalent choices of
coefficients $A_1(x)$, $A_1(x)$. These are:
\[
A_1(x)=3,\quad A_2(x)=-3, \qquad {\rm or} \qquad A_2(x)=A_1(x)+2.
\]
With the first of these choices our final results are that under the change
of variables
\bes
\rho_1(x) & = & {\displaystyle\frac{1}{3}}B_2(x)-Y(x)+\sigma_1(x) \\
\rho_2(x) & = & {\displaystyle\frac{1}{3}}B_2(x)+\sigma_2(x)
\ees
our \Ba\ transformation becomes
\[
y(x)=\sigma_1(x)-\sigma_2(x),
\]
where $\sigma_1(x)$ and $\sigma_2(x)$ satisfy the Riccati system
\bes
\sigma_1'(x) & = & \sigma_1(x)^2+3\sigma_1(x)\sigma_2(x)
                  -{\displaystyle\frac{1}{4}}x \\
\sigma_2'(x) & = & -\sigma_2(x)^2-3\sigma_1(x)\sigma_2(x)
                  +{\displaystyle\frac{1}{4}}x
\ees
and we have in addition the compatibility condition (resulting from the
transformed equation)
\[
\alpha=-\frac{1}{2}
\]
Note that again $Y$ and $a$ play no role in our final results. Eliminating
$\sigma_2(x)$ from the above system and then substituting $\sigma_1(x)=W(x)/2$
yields equation (\ref{ff1}), together with the same B\"ack\-lund transformation 
(\ref{ff2}). On the other hand, eliminating $\sigma_1(x)$ from the above Riccati 
system yields (\ref{ff1}) in $W(x)=-2\sigma_2(x)$.

With the second choice of coefficients $A_i(x)$ we find only that $y(x)=u(x)$
satisfies \PII\ for $\alpha=1/2$, where $u(x)=\rho_1(x)-\rho_2(x)+Y(x)$ is a
solution of
\[
u'(x)=u(x)^2+\frac{1}{2}x
\]
i.e.\ we find only the special integral of \PII\ which gives rise to Airy 
function solutions \cite{gambier:acta}.

\subsection{\PIV }
Dominant balances of the transformed \PIV\ equation yield
only one consistent choice of $A_1$, $A_2$, namely
\[
A_1(x)=A_2(x)=-2.
\]
The resulting equation shows that 
\[
C_2(x)=-2x-2Y(x),\quad B_1(x)=2x+2Y(x),\quad C_1(x)=-4x-2Y(x)+B_2(x),
\]
and that
\bes
\tau_1(x) & = & {\displaystyle\frac{1}{2}} B_2'(x)
               -{\displaystyle\frac{1}{4}}B_2(x)^2
               +xB_2(x)-Y'(x)+Y(x)^2+2xY(x) \\
          & &  -\alpha-1 \pm {\displaystyle\frac{1}{2}} \gamma \\
\tau_2(x) & = & {\displaystyle\frac{1}{2}} B_2'(x) 
               -{\displaystyle\frac{1}{4}}B_2(x)^2
               +xB_2(x)+Y(x)B_2(x)-\alpha-1
               \mp {\displaystyle\frac{1}{2}} \gamma
\ees

Defining $\rho_i$ now by taking
\bes
\rho_1(x)&=&{\displaystyle\frac{1}{2}}B_2(x)-Y(x)-2x+\sigma_1(x)\\
\rho_2(x)&=&{\displaystyle\frac{1}{2}}B_2(x)+\sigma_2(x),
\ees
the BT becomes
\[
y(x)=\sigma_1(x)-\sigma_2(x)-2x.
\]
Once again the intermediate variable $Y(x)$ (and the parameters $a$ 
and $c$) has been eliminated and we are now transforming to $y(x)$ 
from $\sigma_1$ or $\sigma_2$.

Eliminating $\sigma_2(x)$ leads to a different version of \PIV\ 
\[
{\sigma_1}''(x)={{{\sigma_1}'(x)}^2\over 2\sigma_1(x)}
	+{3\over 2}{\sigma_1(x)}^3 - 4x{\sigma_1(x)}^2 	
	+2(x^2-A){\sigma_1(x)}-{C^2\over 2\sigma_1(x)}
\]
where
\bes
A&=&-{1\over 2}\alpha\mp{3\over 4}\gamma-{1\over 2}\\
C^2&=&{1\over 4}(\pm\gamma-2\alpha+2)^2
\ees
(This is the conventional form of \PIV\ if we take $V=-\sigma_1$.)
In this case, the BT becomes
\begin{equation}
y(x) = {{\sigma_1}'(x) + \sigma_1(x)^2 -2x\sigma_1(x) 
        -(1-\alpha\pm\gamma/2) \over 2\sigma_1(x)}.
\end{equation}

Elimination of $\sigma_1(x)$ leads to a conventional version of \PIV\ 
governing $\sigma_2(x)$ but with new $A$ and $C$ given by
\bes
A&=&-{1\over 2}\alpha\pm{3\over 4}\gamma+{1\over 2}\\
C^2&=&{1\over 4}(\pm\gamma+2\alpha+2)^2
\ees
In this case, the BT is
\begin{equation}
y(x) = -{{\sigma_2}'(x) + \sigma_2(x)^2 +2x\sigma_2(x) 
        +(1+\alpha\pm\gamma/2) \over 2\sigma_2(x)}.
\end{equation}

Thus once again we recover the BTs $\widetilde{y}$ and $\hat{y}$ obtained in
Section 3. Considering also the BT from $\sigma_2$ to $\sigma_1$ (respectively 
$\sigma_1$ to $\sigma_2$) then gives, for suitable choices of signs, $\hat{y}=
\widetilde{y}\circ\widetilde{y}$ (respectively $\widetilde{y}=\hat{y}\circ
\hat{y}$) \cite{bch:95}.

For \PIV , the Riccati system (\ref{rr1}), (\ref{rr2}) gives only 
restricted results. We get that $y(x)=u(x)$ is a solution of \PIV\ with
parameters subject to the constraint $\gamma^2=4(\alpha+1)^2$, where 
$u(x)=\rho_1(x)-\rho_2(x)+Y(x)$ satisfies 
\[
u'(x)=u(x)^2+2xu(x)-2(\alpha+1),
\]
i.e.\ we find only a special integral of \PIV .

\section{Conclusions}
We have introduced a new, general method of constructing \Ba\ 
transformations for ordinary differential equations. This is based on
mappings preserving natural subsets of movable poles. For the examples 
considered here this method has allowed us to construct all known fundamental 
\Ba\ transformations, including a less well known \Ba\ transformation for 
\PII . Our approach has also allowed us to find \Ba\ transformations onto 
other ODEs in the Painlev\'e classification, as well as to deduce 
relationships between the \Ba\ transformations constructed. 
Its application to other 	
Painlev\'e equations is discussed in \cite{gjp2}.

\section*{Acknowledgements}

Andrew Pickering and Pilar R.\ Gordoa are grateful to Nalini Joshi 
for 
her invitations to visit the University of Adelaide. The 
research in this paper was supported by the Australian Research Council.


\newpage

\begin{thebibliography}{10}

\bibitem{ac:cup}
M.~J. Ablowitz and P.~A. Clarkson.
\newblock {\em Solitons, Nonlinear Evolution Equations and Inverse Scattering},
  volume 149 of {\em London Mathematical Society Lecture Notes in Mathematics}.
\newblock Cambridge University Press, Cambridge, 1991.

\bibitem{ars:jmp}
M.~J. Ablowitz, A.~Ramani, and H.~Segur.
\newblock A connection between nonlinear evolution equations and ordinary
  differential equations of {P-type} {I} and {II}.
\newblock {\em J. Math. Phys.}, 21:715--721, 1006--1015, 1980.

\bibitem{kjh:96}
M.~D. Kruskal, N.~Joshi, and R.~Halburd.
\newblock Analytic and asymptotic methods for nonlinear singularity analysis: a
  review and extensions of tests for the {Painlev\'e} property.
\newblock In {\em Integrability of Nonlinear Systems}, volume 495 of {\em
  Lecture Notes in Physics}, pages 171--205, Heidelberg, 1997. Springer-Verlag.

\bibitem{mdkpc:pp}
M.~D. Kruskal and P.~A. Clarkson.
\newblock The {Painlev\'e-Kowalevski} and {poly-Painlev\'e} tests for
  integrability.
\newblock {\em Stud. Appl. Math.}, 86:87--165, 1992.

\bibitem{cfp:neg}
R.~Conte, A.~P. Fordy, and A.~Pickering.
\newblock A perturbative {Painlev\'e} approach to nonlinear differential
  equations.
\newblock {\em Physica}, 69D:33--58, 1993.

\bibitem{weiss:II}
J.~Weiss.
\newblock The {Painlev\'e} property for partial differential equations {II}:
  {B\"acklund} transformation, {Lax} pairs, and the {Schwarzian} derivative.
\newblock {\em J. Math. Phys.}, 24:1405--1413, 1983.

\bibitem{wtc:jmp}
J.~Weiss, M.~Tabor, and G.~Carnevale.
\newblock The {Painlev\'e} property for partial differential equations.
\newblock {\em J. Math. Phys.}, 24:522--526, 1983.

\bibitem{njjap:nonlin}
N.~Joshi and J.~A. Petersen.
\newblock A method of proving the convergence of the {Painlev\'e} expansions of
  partial differential equations.
\newblock {\em Nonlinearity}, 7:595--602, 1994.

\bibitem{njgs:nonlin}
N.~Joshi and G.~K. Srinivasan.
\newblock The radius of convergence and well-posedness of the {Painlev\'e}
  expansions of the {Korteweg-deVries} equation.
\newblock {\em Nonlinearity}, 10:71--79, 1997.

\bibitem{egmr93}
P.~G.~Est\'evez, P.~R.~Gordoa, L.~Mart\'\i nez Alonso, and E.~M. Reus.
\newblock Modified singular manifold expansion: application to the {Boussinesq}
  and {Mikhailov}-{Shabat} systems.
\newblock {\em J.~Phys.~A: Math. and General}, 26:1915--1925, 1993.

\bibitem{pilar1}
P.~R.~Gordoa and P.~G.~Est\'evez.
\newblock Double singular manifold method for the {MKDV} equation.
\newblock {\em Teor. Matem. Fizika}, 99:370--376, 1994.

\bibitem{mc:1}
M.~Musette and R.~Conte.
\newblock The two-singular manifold method: {I} modified {KdV} and
  {sine-Gordon} equations.
\newblock {\em J.Phys. A: Math and General}, 27:3895--3913, 1994.

\bibitem{cmp:2}
R.~Conte, M.~Musette, and A.~Pickering.
\newblock The two-singular manifold method: {II} classical {Boussinesq} system.
\newblock {\em J.Phys. A: Math and General}, 28:179--187, 1995.

\bibitem{p:96}
A.~Pickering.
\newblock The singular manifold method revisited.
\newblock {\em J. Math. Phys.}, 37:1894--1927, 1996.

\bibitem{pilar2}
P.~G.~Est\'evez and P.~R.~Gordoa.
\newblock Darboux transformations via {Painlev\'e} analysis.
\newblock {\em Inverse Problems}, 13:939--957, 1997.

\bibitem{w:84a} 
J.~Weiss. \Ba\ transformation and linearizations of the H\'enon-Heiles system.
\newblock {\em Phys. Lett. A}, 102:329-331, 1984.

\bibitem{w:84b}
J.~Weiss. \Ba\ transformation and the H\'enon-Heiles system.
\newblock {\em Phys. Lett. A}, 105:387-389, 1984.

\bibitem{ntz:87} A.~C.~Newell, M.~Tabor, and Y.~B.~Zeng.
\newblock A unified approach to Painlev\'e expansions.
\newblock {\em Physica D}, 29:1-68, 1987.

\bibitem{gntz:88}
J.~D.~Gibbon, A.~C.~Newell, M.~Tabor, and Y.~B.~Zeng.
\newblock Lax pairs, {B\"acklund} transformations and special solutions for
  ordinary differential equations.
\newblock {\em Nonlinearity}, 1:481--490, 1988.

\bibitem{cjp} 
P.~A. Clarkson, Nalini Joshi, and Andrew Pickering.
\newblock {B\"acklund} transformations for the second {Painlev\' e} hierarchy:
  a modified truncation approach.
\newblock {\em Inverse Problems}, 15: 175--187, 1999.

\bibitem{jp} 
N.~Joshi and A.~Pickering. \Ba\ transformations for similarity reductions of
the modified Sawada-Kotera / Kaup-Kupershmidt hierarchy, 
\newblock submitted, 1999.

\bibitem{njmdk:direct}
N.~Joshi and M.~D. Kruskal.
\newblock A direct proof that the solutions of the six {Painlev\'e} equations
  have no movable singularities except poles.
\newblock {\em Stud. Appl. Math.}, 93:187--207, 1994.

\bibitem{gjp2} P.~R.~Gordoa, N.~Joshi and A.~Pickering, 
\newblock Mappings preserving locations of movable poles II,
in preparation, 1999.

\bibitem{gambier:acta}
B.~Gambier.
\newblock Sur les equations differentielles du second ordre et du premier
  degr\'e dont l'int\'egrale g\'en\'erale est \`a points critiques fixes.
\newblock {\em Acta Math.}, 33:1--55, 1910.

\bibitem{bch:95}
A.~Bassom, P.~A. Clarkson, and A.~C. Hicks.
\newblock {B\"acklund} transformations and hierarchies of exact solutions for
  the fourth {Painlev\'e} equation.
\newblock {\em Stud. Appl. Math.}, 95:1--71, 1995.

\bibitem{ince}
E.~L.~Ince.
\newblock {\em Ordinary Differential Equations}.
\newblock Dover, New York, 1954.

\end{thebibliography}
\end{document}